\documentclass[conference]{IEEEtran}
\IEEEoverridecommandlockouts
\usepackage{cite}
\usepackage[per-mode=symbol]{siunitx}
\usepackage{amsmath,amssymb,amsfonts}
\usepackage{graphicx}
\usepackage{bbm} 
\usepackage[nocomma]{optidef}
\usepackage{textcomp}
\usepackage{xcolor}
\usepackage{tabu}
\usepackage{multicol}
\usepackage{lipsum}
\usepackage{tabularx,booktabs}
\usepackage{float}
\usepackage{array}
\newcolumntype{C}{>{\centering\arraybackslash}X}
\usepackage{wrapfig}
\usepackage{multirow}
\usepackage{tabularx}
\usepackage{pgfplots}
\usepackage[font=scriptsize,skip=0pt]{caption}
\usepackage{subcaption}
\usepackage{float}
\usepackage{pgfplotstable}
\usepackage{siunitx}
\usepackage{tikz}
\usepackage{adjustbox}

\usepackage{booktabs}
\usepackage{pifont}
\usepackage[linesnumbered,ruled,vlined]{algorithm2e}
    
\SetCommentSty{mycommfont}
\SetKwInput{KwInput}{Input}
\SetKwInput{KwOutput}{Output}
\SetKwInput{KwRequire}{Require}
\pgfplotsset{compat=1.18}
\def\BibTeX{{\rm B\kern-.05em{\sc i\kern-.025em b}\kern-.08em
    T\kern-.1667em\lower.7ex\hbox{E}\kern-.125emX}}

    \usepackage{acronym}    
    \acrodef{ac}[AC]{Admission Control}
    \acrodef{acm}[ACM]{Admission Control Module}
    \acrodef{acr}[AR]{Acceptance Ratio}
    \acrodef{aru}[ARU]{Average Resource Utilization}
    \acrodef{aco}[ACO]{Ant Colony Optimization}
    \acrodef{adc}[AC]{Admission Control}
    \acrodef{adcpe}[ACP]{Admission Control Policies}
    \acrodef{adcp}[ACP]{Admission Control Policy}
    \acrodef{af}[AF]{Application Function}
    \acrodef{api}[API]{Application Programming Interface}
    \acrodef{aoi}[AoI]{Age of Information}
    \acrodef{apollo}[APOLLO]{bAndit-based oPtimization Of muLtipLe Objectives}
    \acrodef{ar}[AR]{Augmented Reality}
    \acrodef{arima}[ARIMA]{Auto-regressive Integrated Moving Average}
    \acrodef{awgn}[AWGN]{Additive White Gaussian Noise}
    \acrodef{be}[BE]{Best Effort}
    \acrodef{bs}[BS]{Base Station}
    \acrodef{bl}[BL]{Bandit Learning}
    \acrodef{bw}[BW]{Bandwidth}
    \acrodef{bbu}[BBU]{Baseband Unit}
    \acrodef{bnn}[BNN]{Bayesian Neural Network}
    \acrodef{bpsk}[BPSK]{Binary Phase Shift Keying}
    \acrodef{c2ucb}[C$^2$UCB]{Contextual-Combinatorial Upper-Confidence Bound}
    \acrodef{ca}[CA]{Cluster Agent}
    \acrodef{capex}[CAPEX]{Capital Expenditure}
    \acrodef{camra}[CAMRA]{Context-Aware Multi Resource Allocation}
    \acrodef{cc}[CC]{Cloud Computing}
    \acrodef{ceu}[CEU]{Central Edge Unit}
    \acrodef{comoucb}[COMO-UCB]{Combinatorial Multi-Objective Upper-Confidence Bound}
    \acrodef{comomab}[COMO-MAB]{Combinatorial Multi-Objective Multi-Armed Bandit}
    \acrodef{cmab}[C-MAB]{Contextual Multi-Armed Bandit}
    \acrodef{cb}[CB]{Contextual Bandit}
    \acrodef{cn}[CN]{Core Network}
    \acrodef{cnn}[CNN]{Convolutional Neural Network}
    \acrodef{cnf}[CNF]{Cloud Native Function}
    \acrodef{cns}[CNS]{Cloud Network Slicing}
    \acrodef{cpu}[CPU]{Central Processing Unit}
    \acrodef{ctr}[CTR]{Click Through Rate}
    \acrodef{cts}[CTS]{Contextual Thompson Sampling}
    \acrodef{csp}[CSP]{Communication Service Provider}
    \acrodef{cu}[CU]{Central Unit}
    \acrodef{dag}[DAG]{Directed Acyclic Graph}
    \acrodef{dc}[DC]{Data Center}
    \acrodef{dp}[DP]{Dynamic Programming}
    \acrodef{dl}[DL]{Deep Learning}
    \acrodef{dlrl}[(D)RL]{(Deep) Reinforcement Learning}
    \acrodef{dil}[DL]{Distributed Learning}
    \acrodef{drl}[DRL]{Deep Reinforcement Learning}
    \acrodef{dnn}[DNN]{Deep Neural Network}
    \acrodef{dt}[DT]{Digital Twin}
    \acrodef{dt2}[DT2]{Deutsche Telekom} 
    \acrodef{du}[DU]{Distributed Unit}
    \acrodef{e2e}[E2E]{End-to-End}
    \acrodef{ec}[EC]{Edge Computing}
    \acrodef{embb}[eMBB]{enhanced Mobile Broadband}
    \acrodef{en}[EN]{Edge Node}
    \acrodef{enw}[EN]{Edge Network}
    \acrodef{ens}[ENS]{Edge Network Slicing}
    \acrodef{epc}[EPC]{Evolved Packet Core}
    \acrodef{es}[ES]{Edge Server}
    \acrodef{esp}[ESP]{Edge Slice Provisioning}
    \acrodef{etsi}[ETSI]{European Telecommunications Standards Institute}
    \acrodef{exprp}[ExpRP]{Exponential Reservation Policy}
    \acrodef{faas}[FaaS]{Function-as-a-Service}
    \acrodef{fcfs}[FCFS]{First Come First Serve}
    \acrodef{fh}[FH]{Fronthaul}
    \acrodef{fip}[FIP]{Fog Infrastructure Provider}
    \acrodef{fn}[FN]{Fog Node}
    \acrodef{fsp}[FSP]{Fog Service Provisioning}    
    \acrodef{fomdmkp}[FOMDMKP]{Fractional Online  Multi-Dimensional Multiple Knapsack Problem}
    \acrodef{fpga}[FPGA]{Field-Programmable Gate Array}
    \acrodef{ftl}[FTL]{Follow-The-Leader}
    \acrodef{ftpl}[FTPL]{Follow-The-Perturbed-Leader}
    \acrodef{ga}[GA]{Genetic Algorithm}
    \acrodef{gpp}[GPP]{General Purpose Processor}
    \acrodef{gppj}[GPP]{Generation Partnership}
    \acrodef{gpu}[GPU]{Graphic Processor Unit}
    \acrodef{gsm}[GSM]{Global System for Mobile Communications}
    \acrodef{ggi}[GGI]{Generalized Gini Index}
    \acrodef{ggf}[GGF]{Generalized Gini Function}
    \acrodef{ggwf}[GGWF]{Generalized Gini Welfare Function}
    \acrodef{helios}[HELIOS]{\underline{H}ierarchical n\underline{E}twork s\underline{L}\underline{I}ce pr\underline{O}vi\underline{S}ioning}
    \acrodef{hb}[HB]{Hierarchical Bandit}
    \acrodef{hmab}[HMAB]{Hierarchical Multi-Armed Bandit}
    \acrodef{hdf}[HDF]{Hierarchical Decision Framework}    
    \acrodef{hla}[HLA]{High-Level Agent}
    \acrodef{hrl}[HRL]{Hierarchical Reinforcement Learning}
    \acrodef{hdrl}[HDRL]{Hierarchical Deep Reinforcement Learning}
    \acrodef{lla}[LLA]{Low-Level Agent}
    \acrodef{gt}[GT]{Game Theory}
    \acrodef{ip}[IP]{Integer Program}
    \acrodef{ilp}[ILP]{Integer Linear Program}
    \acrodef{inp}[InP]{Infrastructure Provider}
    \acrodef{iot}[IoT]{Internet-of-Things}
    \acrodef{itu}[ITU]{International Telecommunications Union}
    \acrodef{kpi}[KPI]{Key Performance Indicators}
    \acrodef{kf}[KF]{Kalman Filter}
    \acrodef{lp}[LP]{Linear Program}
    \acrodef{linrp}[LinRP]{Linear Reservation Policy}
    \acrodef{lmgtfy}[LMGTFY]{Let Me Google That For You}
    \acrodef{linucb}[LinUCB]{Linear Upper Confidence Bound}
    \acrodef{lxc}[LXC]{Linux Container}
    \acrodef{mab}[MAB]{Multi-Armed Bandit}
    \acrodef{marl}[MARL]{Multi-Agent Reinforcement Learning}
    \acrodef{mabp}[MABP]{Multi-Armed Bandit Problem}
    \acrodef{mac}[MAC]{Medium Access Protocol}
    \acrodef{mano}[MANO]{Management \& Orchestration}
    \acrodef{mcs}[MCS]{Modulation and Coding Scheme}
    \acrodef{mdp}[MDP]{Markov Decision Process}
    \acrodef{mdkp}[MdKP]{Multidimensional Knapsack Problem}
    \acrodef{mec}[MEC]{Mobile Edge Computing}
    \acrodef{mimo}[MIMO]{Multiple Input Multiple Output}
    \acrodef{milp}[MILP]{Multiple Integer Linear Program}
    \acrodef{madrl}[MADRL]{Multi-Agent Deep Reinforcement Learning}
    \acrodef{momab}[MOMAB]{Multi-Objective Multi-Armed Bandit}
    \acrodef{molinucb}[MOLinUCB]{Multiple-Objective Linear Upper-Confidence Bound}
    \acrodef{mobo}[MOBO]{Multi-Objective Bandit Optimization}
    \acrodef{moba}[MOBA]{Multi-Objective Bandit Algorithm}
    \acrodef{mocmab}[MO-CMAB]{Multi-Objective Contextual Multi-Armed Bandit}
    \acrodef{mo}[MO]{Multiple-Objective}
    \acrodef{moo}[MOO]{Multi-Objective Optimization}
    \acrodef{moop}[MOOP]{Multi-Objective Optimization Problem}
    \acrodef{moonline}[MOONLINE]{\underline{M}ulti-\underline{O}bjective bandit model for \underline{ONLINE}}
    \acrodef{ml}[ML]{Machine Learning}
    \acrodef{mlops}[MLOps]{Machine Learning Operations}
    \acrodef{mm}[MM]{Markov Model}
    \acrodef{mmtc}[mMTC]{massive Machine Type Communication}
    \acrodef{mno}[MNO]{Mobile Network Operator}
    \acrodef{naas}[NaaS]{Network-as-a-Service}
    \acrodef{nf}[NF]{Network Function}
    \acrodef{nfv}[NFV]{Network Function Virtualization}
    \acrodef{ngmn}[NGMN]{Next-Generation Mobile Network}
    \acrodef{ngmna}[NGMNA]{Next-Generation Mobile Networks Alliance}
    \acrodef{nn}[NN]{Neural Network}
    \acrodef{nr}[NR]{New Radio}
    \acrodef{nrb}[NRB]{Network Resource Broker}
    \acrodef{ns}[NS]{Network Slicing}
    \acrodef{nsl}[NSL]{Network Slice}
    \acrodef{nsb}[NSB]{Network Slice Broker}
    \acrodef{nse}[NSE]{Network Slice Embedding}
    \acrodef{nsp}[NSP]{Network Slice Provider}
    \acrodef{nspr}[NSP]{Network Slice Provisioning}
    \acrodef{nspl}[NSP]{Network Slice Placement}
    \acrodef{nst}[NST]{Network Slice Template}
    \acrodef{nsr}[NSR]{Network Slice Request}
    \acrodef{nsrp}[NSRP]{Network Slice Request Placement}
    \acrodef{nt}[NT]{Network Topology}
    \acrodef{oco}[OCO]{Online Convex Optimization}
    \acrodef{ogd}[OGD]{Online Gradient Descent}
    \acrodef{oga}[OGA]{Online Gradient Ascent}
    \acrodef{ol}[OL]{Online Learning}
    \acrodef{oml}[OML]{Online Meta-Learning}
    \acrodef{osacp}[OSACP]{Online Slice Admission Control Problem}
    \acrodef{osaca}[OSACA]{Online Slice Admission Control-Adaptive}
    \acrodef{osp}[OSP]{Online Slice Placement}
    \acrodef{oens}[OENS]{Online Edge Node Selection}
    \acrodef{ofns}[OFNS]{Online Fog Node Selection}
    \acrodef{ofdma}[OFDMA]{Orthogonal Frequency-Division Multiple Access}
    \acrodef{omdkp}[OMdKP]{Online Multidimensional Knapsack Problem}
    \acrodef{omdmkp}[OMdMKP]{Online Multi-dimensional Multiple Knapsack Problem}
    \acrodef{onsp}[ONSP]{Online Network Slice Provisioning}
    \acrodef{okp}[OKP]{Online Knapsack Problem}
    \acrodef{opex}[OPEX]{Operating Expenditure}
    \acrodef{osac}[OSAC]{Online Slice Admission Control}
    \acrodef{osp}[OSP]{Online Slice Placement}
    \acrodef{oran}[O-RAN]{Open Radio Access Network}
    \acrodef{owa}[OWA]{Ordered Weighted Averaging}
    \acrodef{pf}[PF]{Pareto Front}
    \acrodef{pg}[PG]{Parallel Graph}    
    \acrodef{pdcp}[PDCP]{Packet Data Convergence Protocol}
    \acrodef{pnf}[PNF]{Physical Network Function}
    \acrodef{prb}[PRB]{Physical Resource Block}
    \acrodef{qam}[QAM]{Quadrature Amplitude Modulation}
    \acrodef{qoe}[QoE]{Quality of Experience}
    \acrodef{qos}[QoS]{Quality of Service}
    \acrodef{qpsk}[QPSK]{Quadrature Phase Shift Keying}
    \acrodef{ra}[RA]{Resource Allocation}
    \acrodef{rac}[RAC]{Radio Admission Control}
    \acrodef{ram}[RAM]{Random Access Memory}
    \acrodef{ran}[RAN]{Radio Access Network}
    \acrodef{rans}[RaNS]{Radio Network Slicing}
    \acrodef{resb}[ResB]{Resource Broker}
    \acrodef{rb}[RB]{Resource Block}
    \acrodef{rl}[RL]{Reinforcement Learning}
    \acrodef{rnt}[RNT]{Real Network Topology}
    \acrodef{rlc}[RLC]{Radio Link Control}
    \acrodef{rrc}[RRC]{Radio Resource Control}
    \acrodef{rrm}[RRM]{Radio Resource Management}
    \acrodef{ru}[RU]{Radio Unit}
    \acrodef{se}[SE]{Spectrum Efficiency}
    \acrodef{sac}[SAC]{Slice Admission Control}        
    \acrodef{sec}[SEC]{Serverless Edge Computing}    
    \acrodef{sacps}[SACPS]{Slice Admission Control Policy Selection}    
    \acrodef{sacp}[SACP]{Slice Admission Control Policy}    
    \acrodef{sba}[SBA]{Service-Based Architecture}
    \acrodef{scfdma}[SCFDMA]{Single Carrier-Frequency-Division Multiple Access}
    \acrodef{sdc}[SDC]{Software Defined Controller}    
    \acrodef{sefc}[SEFC]{Serverless Functions Chain}
    \acrodef{sdn}[SDN]{Software Defined Networking}
    \acrodef{sdnc}[SDNC]{Software Defined Network Controller}
    \acrodef{sdr}[SDR]{Software Defined Radio}
    \acrodef{sf}[SF]{Serverless Functions}
    \acrodef{sfc}[SFC]{Service Function Chain}
    \acrodef{sfcp}[SFCP]{Service Function Chain Provisioning}
    \acrodef{siso}[SISO]{Single Output Single Input}
    \acrodef{sla}[SLA]{Service Level Agreement}    
    \acrodef{soa}[SoA]{State-of-the-Art}
    \acrodef{soco}[SOCO]{Online Convex Optimization with Switching Cost}
    \acrodef{spr}[SP]{Service Provider}
    \acrodef{sr}[SR]{Service Request}
    \acrodef{sla}[SLA]{Service Level Agreement}
    \acrodef{slo}[SLO]{Service Level Objective}
    \acrodef{slaas}[SlaaS]{Slice-as-a-Service}
    \acrodef{slr}[SLR]{Slice Request}
    \acrodef{so}[SO]{Slice Orchestrator}
    \acrodef{soo}[SOO]{Single Objective Optimization}
    \acrodef{sor}[SO]{Slice Orchestration}
    \acrodef{sp}[SP]{Slice Placement}
    \acrodef{st}[ST]{Slice Tenant}
    \acrodef{sota}[SoTA]{State-of-The-Art}
    \acrodef{sst}[SST]{Slice/Service Type}
    \acrodef{sw}[SW]{Sliding Window}
    \acrodef{tn}[TN]{Transport Network}
    \acrodef{ucb}[UCB]{Upper Confidence Bound}
    \acrodef{ue}[UE]{User Equipment}
    \acrodef{ul}[UL]{Uplink}
    \acrodef{urllc}[uRLLC]{ultra-Reliable and Low-Latency Communication}
    \acrodef{v2x}[V2X]{Vehicle-to-Everything}
    \acrodef{vec}[VEC]{Virtual Edge Computing}
    \acrodef{vim}[VIM]{Virtualized Infrastructure Manager}
    \acrodef{vm}[VM]{Virtual Machine}
    \acrodef{vne}[VNE]{Virtual Network Embedding}
    \acrodef{vnf}[VNF]{Virtual Network Function} 
    \acrodef{vnffg}[VNF-FG]{Virtual Network Function Forwarding Graph} 
    \acrodef{vnfpg}[VNF-PG]{Virtual Network Function Parallel Graph} 
    \acrodef{vnr}[VNR]{Virtual Network Request}
    \acrodef{vr}[VR]{Virtual Reality}
    \acrodef{vl}[VL]{Virtual Link}
    \acrodef{vran}[vRAN]{Virtual Radio Access Network}
    \acrodef{wtp}[WTP]{Willingness To Pay}
    \acrodef{wtpr}[WTPR]{Willingness-To-Pay-Ratio}

\begin{document}

\title{Hierarchical Placement Learning\\ for Network Slice Provisioning
}

\author{\IEEEauthorblockN{Jesutofunmi Ajayi}
\IEEEauthorblockA{\textit{Institute of Computer Science} \\
\textit{University of Bern}\\
Bern, Switzerland \\
\{firstname.lastname\}@unibe.ch}
\and
\IEEEauthorblockN{Antonio Di Maio}
\IEEEauthorblockA{\textit{Institute of Computer Science} \\
\textit{University of Bern}\\
Bern, Switzerland \\
\{firstname.lastname\}@unibe.ch}
\and
\IEEEauthorblockN{Torsten Braun}
\IEEEauthorblockA{\textit{Institute of Computer Science} \\
\textit{University of Bern}\\
Bern, Switzerland \\
\{firstname.lastname\}@unibe.ch}
}

\maketitle

\begin{abstract}    
    In this work, we aim to address the challenge of slice provisioning in edge-based mobile networks.
    We propose a solution that learns a service function chain placement policy for \aclp{nsr}, to maximize the request acceptance rate, while minimizing the average node resource utilization.
    To do this, we consider a \acl{hmab} problem and propose a two-level hierarchical bandit solution which aims to learn a scalable placement policy that optimizes the stated objectives in an online manner.     
    Simulations on two real network topologies show that our proposed approach achieves 5\% average node resource utilization while admitting over 25\% more slice requests in certain scenarios, compared to baseline methods. 
\end{abstract}

\begin{IEEEkeywords}
Network Slicing, Multi-Objective Optimization, \acl{ol}, Edge Networks
\end{IEEEkeywords}

\section{Introduction} \label{sec:intro}
    
    In next generation mobile networks, service providers will be expected to support complex, and increasingly heterogeneous service requirements~\cite{10090468}.
    By combining \ac{sdn} and \ac{nfv} to provide \ac{ns},  which is a novel virtualized infrastructure model~\cite{9627736} in 5G and beyond mobile networks, service providers will be able to meet these requirements while concurrently provisioning and multiplexing a diverse set of services over a shared communication infrastructure.
    Through \ac{ns}, multiple dedicated and virtualized networks can be configured to meet a diverse range of service requirements.        

    The effective management of network resources in next generation mobile networks poses a significant challenge due to the dynamic nature and size of the network environment, especially at the network edge.
    This challenge motivates the development of intelligent and scalable solutions that can be used to address the network slice provisioning problem, which relates to determining the locations where the \acp{vnf} of a service function chain, will be deployed~\cite{10625023}.    
    The majority of solutions to the network slice provisioning problem are typically designed for the offline setting or leverage a single-agent in the online setting, which typically leads to them under performing in dynamic, large-scale networks.
    Moreover, the current network slice provisioning solutions aim to address a single provisioning objective, leading to inefficient resource utilization, limited flexibility, and the inability to simultaneously satisfy diverse service requirements.
    
    Therefore, in this work, we address the following research questions:       
    \begin{itemize}
        \item \textit{RQ1}: How to increase the rate of admitted \acp{nsl} while reducing the required allocated resources, especially in \textbf{large-scale} networks with highly \textbf{dynamic} resource and user requirements?
        \item \textit{RQ2}: How can a \textbf{hierarchical model} be devised so that agents in different network subdomains make their own placement decisions?
        \item \textit{RQ3}: How to design a \ac{nsl} provisioning performance metric that considers \textbf{multiple objectives} and enables a network policymaker to specify the objectives' relative \textbf{importance} and mutual \textbf{fairness}?
    \end{itemize}
    
    To address these research questions, we make the following contributions towards the design of an intelligent and scalable network slice provisioning solution:
    
    \begin{enumerate}

        \item 
        We formulate the network slice provisioning problem as a general offline constrained optimization problem, considering computing and network resources as constraints.
        To allow an online solution, we reformulate network slice provisioning problem as a \ac{hmab} problem.        
        
        \item
        We propose \ac{helios}, a two-level hierarchical learning algorithm to solve the online network slice provisioning problem. 
        At the high-level, a contextual bandit agent directs each slice request to a specific cluster region in the network, depending on the measured resource state and slice features.
        At the low-level, a combinatorial bandit agent determines the nodes on which the \acp{vnf} will be placed. 
        \ac{helios} is designed to learn a placement policy that scales with network size. 

        \item 
        We leverage the \ac{ggi} aggregation function which scalarizes and balances multiple provisioning objectives, together.
        By maximizing this function, we aim to find a point on the Pareto front of the multi-objective optimization problem.
        
    \end{enumerate}

    Through extensive simulations, we evaluate the performance of our approach in terms of the \acl{aru}, and Average \acl{acr}, compared to various baseline solutions, on real network topologies. 

    The remainder of this paper is organized as follows. 
    In the next section, we present the related work. 
    We describe the system model in Section~\ref{sec:systemodel} and formulate the network slice provisioning  problem.
    Then, in Section~\ref{sec:MOONline}, we consider the problem in the bandit setting and proposed our solution, \ac{helios}.
    In Section~\ref{sec:perfeval}, we evaluate our proposed approach compared to other baseline solutions.    
    Finally, we conclude the paper in Section~\ref{sec:conclusion}.

    \section{Related Work} \label{sec:relatedwork}
    Toumi et al. propose a \ac{drl} approach for \ac{sfc} placement in multiple domains~\cite{toumi2021hierarchical}. 
    Specifically, they propose a hierarchical architecture where the agents in local domain and a multi-domain agent are trained using different \ac{drl} models to place \acp{sfc}.
    Mao et al. propose a meta-heuristic algorithm for the online deployment of \acp{sfc}, where they seek to jointly minimize the server operation cost as well as the network latency~\cite{ACOL}.     
    Wu et al. address the network slice provisioning problem by proposing an \ac{rl}-based framework that seeks to maximize the long-term revenue of network operators through slice admission and resource allocation~\cite{10278745}.      
    Tran et al. propose a \ac{drl} framework for 5G \ac{sfc} provisioning, where \ac{drl} is used for joint \ac{sfc} placement and routing~\cite{10623413}.         
    Finally, Qiao et al. propose a hierarchical \ac{drl}-based algorithm to address the resource allocation problem in \ac{oran} network slicing~\cite{qiao2025resource}.
    
    Previous work on network slice provisioning has limitations: centralized approaches struggle with scalability and require full system knowledge, while decentralized ones, though more scalable, lack sufficient coordination. Hybrid solutions with hierarchical multi-agent DRL aim to balance these issues but are inefficient in dynamic settings due to high training costs and slow convergence~\cite{9903386}. 
    Our method instead uses a hierarchical \ac{mab} framework, which is lightweight, sample-efficient, and better suited to dynamic environments thanks to fast, adaptive decision-making~\cite{8421637}. 
    To the best of our knowledge, this is the first hierarchical \ac{mab} solution proposed for network slice provisioning in distributed edge networks.
    
    \section{System Model and Problem Formulation} \label{sec:systemodel}
    
    \subsection{System Model}
    \begin{figure}
      \includegraphics[ width=\columnwidth]{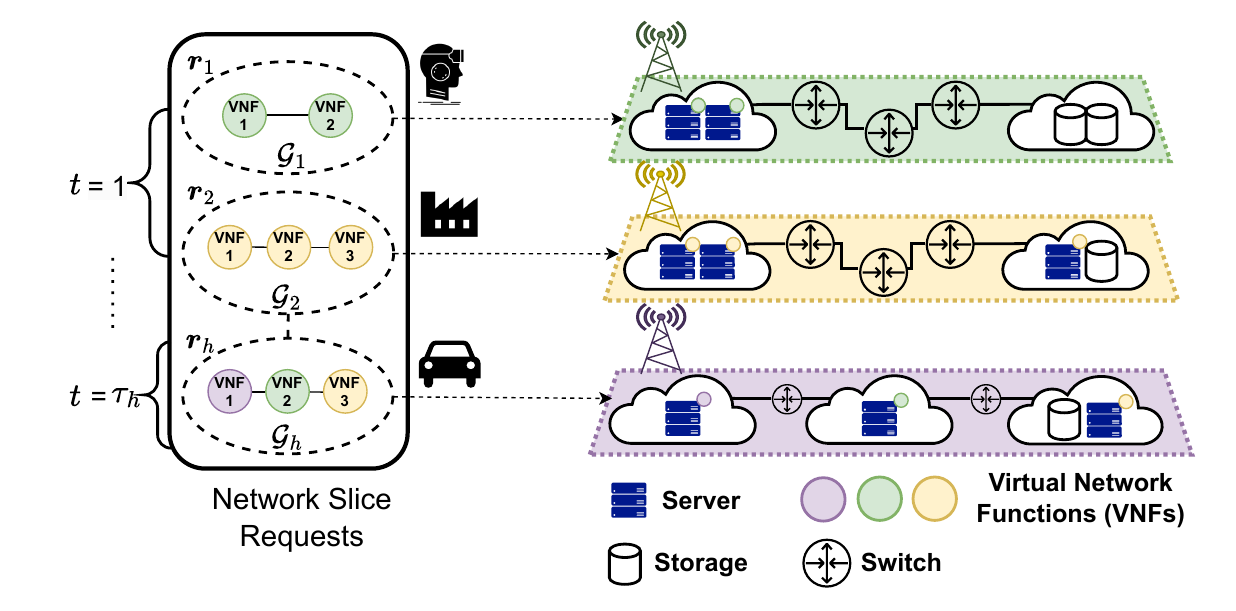}
      \caption{System Model}
      \label{fig:Architecture}  
    \end{figure}
    
    We consider a distributed edge-computing network primarily comprising of edge servers and physical paths that are required to support the provisioning of diverse \acp{nsr}~\cite{10480301}, see Fig~\ref{fig:Architecture}.
    Each edge server has an arbitrary but limited amount of compute and network resources.
    Such a network can be modeled as a connected, undirected graph $G$ = ($N$, $E$), where $N$ represents the set of edge servers available in the network and $E$ represents the set of physical links in the network.
    In the considered network scenario, we assume the availability of node-specific resources such as CPU (MIPS), GPU (GFLOPS), RAM ($\si{\mega\byte}$), and Storage ($\si{\mega\byte}$), as well as link-specific resources like Bandwidth ($\si{\mega\bit\per\second}$) for communication between servers.
    We define $C^n_j$ as the maximum capacity of resource type $j$ on a network node $n \in N$, and $C^q_j$ as the maximum capacity of resource type $j$ on a network link $q \in E$.    
    To avoid ambiguity when aggregating these values, we define the total capacity for each category separately. 
    The aggregated capacity for the $j$-th type of node resource across all nodes is given by $C^{\text{node}}_j = \sum_{n \in N} C^n_j$. 
    Similarly, the aggregated capacity for the $j$-th type of link resource across all links is given by $C^{\text{link}}_j = \sum_{q \in E} C^q_j$.
    Based on this infrastructure model, we assume that the \ac{inp} owns and leases the physical network resources in an elastic, pay-as-you-go model by dynamically allocating and de-allocating them to incoming and existing \ac{nsr}.        
    In line with the time-varying conditions at the network edge, we model the time in the system as divided into discrete consecutive intervals, i.e., \textit{time slots} $t \in T$. 
    
    In our request model, an \ac{nsr} is identified by the incremental index $h\in \mathcal{H}$, where $\mathcal{H} = \{1, \ldots, |\mathcal{H}|\}$ is defined as the ordered set of all request indices.    
    We define $\boldsymbol{r}_h=(r_{h}^1,\ldots,r_{h}^m, \delta_h)$ as the vector containing the required amount of each resource request, as well as the \textit{lifetime} of the request, $\delta_h \in \mathbb{N}$, which is defined as the number of time slots the request needs to access the network resources.
    We assume that slice tenants or service providers submit requests for \acp{nsl}, where each request defines the resources required by the \ac{nsl} and the duration of the request.    
    Based on the resource requirements of request $h$ defined by $\boldsymbol{r}_h$, the tenant defines a suitable set of \acp{vnf} to provide the network slice request with a sufficient performance level to support the associated application.
    This set of \acp{vnf}, connected together through virtual links, form a \acl{sfc} (\ac{sfc}) which would need to be deployed over the edge network to provision a \ac{nsl} for services with specific performance requirements, as seen in Figure~\ref{fig:Architecture}.    
    
    We indicate the single \ac{sfc} associated with \ac{nsr} $h$ as $\mathcal{G}_h = (N_h, E_h, \delta_h, \tau_h)$, where $N_h$ represents the set of all \acp{vnf} in SFC $G_h$, the quantity $E_h$ represents the set of links concatenating the \acp{vnf} in the request, and $\tau_h \in \mathbb{N}$ represents the \textit{timestamp} of the request, defined as the time slot index at which the request arrives in the network.  

    We denote the multi-dimensional resource requirements of a \ac{vnf} $v$ in a given \ac{sfc} as $\boldsymbol{\varphi}_{v}=(\varphi^1_{v},\ldots,\varphi^{m_{\text{node}}}_{v})$, with $r_h^j=\sum_{v\in N_h} \varphi_v^j$, where $m_\text{node}$ is the number of node resource types.
    We denote the multi-dimensional resource requirements of a virtual link $e =(v, v') \in E_h \subseteq N_h^2$ between a pair of \acp{vnf} $v$ and $v'$ in a given \ac{sfc} $h$ as $\boldsymbol{\varphi}_e=(\varphi^1_e,\ldots,\varphi^{m_{\text{link}}}_e)$, with $r_h^j=\sum_{e\in E_h} \varphi_e^j$, where $m_\text{link}$ is the number of link resource types.    

    \subsection{Problem Formulation} \label{sec:ProbForm}
    
    Given an arriving request for an \ac{sfc}, the objective in the network slice provisioning problem is to determine the placement for each \ac{vnf} in the \ac{sfc}, while jointly achieving the objectives of minimizing average resource utilization and maximizing the number of accepted requests.
    More specifically, given an \ac{sfc} $\mathcal{G}_h$, our goal is to find a strategy that maps each \ac{vnf} $v$ in $N_h$ to an edge server $n \in N$ in the network topology. 

    \subsubsection{Constraints}
    
    We introduce two binary variables $x^n_v$ and $y^q_e$. 
    In the following constraint~(\ref{eq:eq1}), $x^n_v$ indicates a mapping of a \ac{vnf} $v$ to an \ac{en} $n$ at time slot $t$,
    \begin{equation}
        \label{eq:eq1}
        x^n_{{v}} =
        \begin{cases}
          1, & \text{if \ac{vnf} $v$ is placed on \ac{en} $n$}\  \\
          0, & \text{otherwise} \\
        \end{cases}\textbf{}
    \end{equation}

    In constraint~(\ref{eq:eq2}), $y^q_e(t)$ indicates whether a \ac{vl} $e\in E_h$ between two \acp{vnf}, is mapped to a corresponding physical link in the edge network $q\in E$    
    \begin{equation}
        \label{eq:eq2}
        y^q_{e} =
        \begin{cases}    
          1, & \text{if \ac{vl} $e$ is mapped to physical link $q$}\  \\ 
          0, & \text{otherwise.} \\
        \end{cases}\textbf{}
    \end{equation}
    
    We now introduce two families of constraints (\ref{eq:con3}) and (\ref{eq:con4}) to ensure that the nodes' and links' capacities, respectively, are sufficient to support the resource requirements of the arriving \ac{sfc}.    
    \begin{equation}
        \begin{aligned}
            \sum\limits_{h \in \mathcal{H}}\sum\limits_{v \in N_h}  \varphi^j_v x^n_v \mathbf{1}_{[\tau_h \leq t < \tau_h+\delta_h]} \leq C^n_j,  \qquad \, \forall n \in {N}, \\ \,\forall j\in \{1,\ldots,m_\text{node}\}, \,\forall t \in T  \label{eq:con3}
        \end{aligned}
    \end{equation}    
    
    \begin{equation}
        \begin{aligned}
        \sum\limits_{h \in \mathcal{H}}\sum\limits_{e \in E_h} 
        \varphi^j_e y^q_e \mathbf{1}_{[\tau_h \leq t < \tau_h+\delta_h]} \leq C^q_j, \qquad \, \forall {q} \in {E}, \\ \,\forall j\in \{m_\text{node}+1,\ldots,m_\text{node}+m_\text{link}=m\} ,\,\forall t \in T, \label{eq:con4}                    
        \end{aligned}        
    \end{equation}
   
    To ensure that each \ac{vnf} $v_h$ of an \ac{nsr} is mapped to at most one node in the physical network graph $\mathcal{G}$, we define the following constraint (\ref{eq:con5}).

    \begin{equation}
        \label{eq:con5}        
        \sum\limits_{n \in N} x^n_{v} = 1,\,\qquad \forall h \in \mathcal{H}, \forall v \in N_h
    \end{equation}
    
    \subsubsection{Objectives}
    
    The network slice provisioning optimization problem aims to determine, at each time slot, the optimal placement of an \ac{nsr} that satisfies the resource constraints and minimizes a cost function $U$ to achieve a tradeoff between the two following objectives.
    \begin{itemize}
        \item \textit{Maximize Accepted \acp{nsr}}: The objective in (\ref{eq:f1}) aims to maximize the number of accepted requests in the network, which would maximize the revenue gained by the \ac{inp}. 
        Formally, we represent this objective as:
        \begin{equation}
            \label{eq:f1}            
            f_\text{a} = \sum\limits_{n \in N} \sum\limits_{v \in N_h} x^n_v
        \end{equation}
    
        \item \textit{Minimize Resource Utilization}:
        In the second objective, we aim to minimize the average resource utilization of edge nodes. 
        We define the optimization objective for the $j$-th node resource as in Eq. (\ref{eqn:mru_n}).
        
        \begin{equation}
            \label{eqn:mru_n}                
        f_j = - \sum_{n\in N} \sum_{h \in \mathcal{H}} \sum_{v \in N_h}  \varphi^j_v x^n_v, \qquad \forall j\in \{1,\ldots,m_\text{node}\}
        \end{equation} 

        and the optimization objective for the $j$-th type of link resource as in Eq. (\ref{eqn:mru_l}).
        \begin{equation}
        \begin{aligned}            
            f_j = - \sum_{q\in E} \sum_{h \in \mathcal{H}} \sum_{e \in E_h}  \varphi^j_e y^q_e, 
            \qquad \\ \, \forall j\in \{m_\text{node}+1,\ldots,m_\text{node}+m_\text{link}=m\}
        \end{aligned}
        \label{eqn:mru_l}        
        \end{equation} 
        
        \end{itemize}

    \textit{Multi-Objective Optimization}: Let us define $\boldsymbol{\mathbf{w}}=(w_\text{a},w_1,\ldots,w_m)\in[0,1]^{m+1}$ as the weighting parameters for the different objectives, where $\lVert \boldsymbol{\mathbf{w}}\lVert=1$.
    We define the multi-objective utility vector as $\boldsymbol{f}=(f_\text{a},f_1,\ldots,f_m)^\top$.
    We can combine the optimization objectives, while considering the constraints, to formulate a scalarized multi-objective optimization problem as Eq.~(9):
    \begin{maxi!}|s|[2]                 
            {}                          
            {U(\boldsymbol{f})=\boldsymbol{\mathbf{w}}^\top \boldsymbol{f} = w_a f_a + \sum_{j=1}^{m} w_j f_j \label{eq:moop}}
            {}             
            {}             
            \addConstraint (1) - (5)                
    \end{maxi!}  
    
    The above constrained combinatorial optimization problem is generally considered NP-hard~\cite{hu2023giph, li2023alpaserve}, as it would require prior information about the request resource requirements and the performance of \acp{vnf} across various nodes in the network, which might not be readily available or might be computationally expensive to obtain in polynomial time.
    Furthermore, the space of possible placement combinations grows exponentially with the number of edge devices and the length of \acp{sfc}, and while commercial solvers have previously been used to address similar optimization problems \cite{10.1145/3629136}, they assume perfect system knowledge, \textcolor{black}{have a long execution time, and do not necessarily scale well with larger optimization problems~\cite{xu2024zeal}}.
    
    \section{Hierarchical Placement Learning} \label{sec:MOONline}
    \begin{figure}
      \includegraphics[ width=\columnwidth]{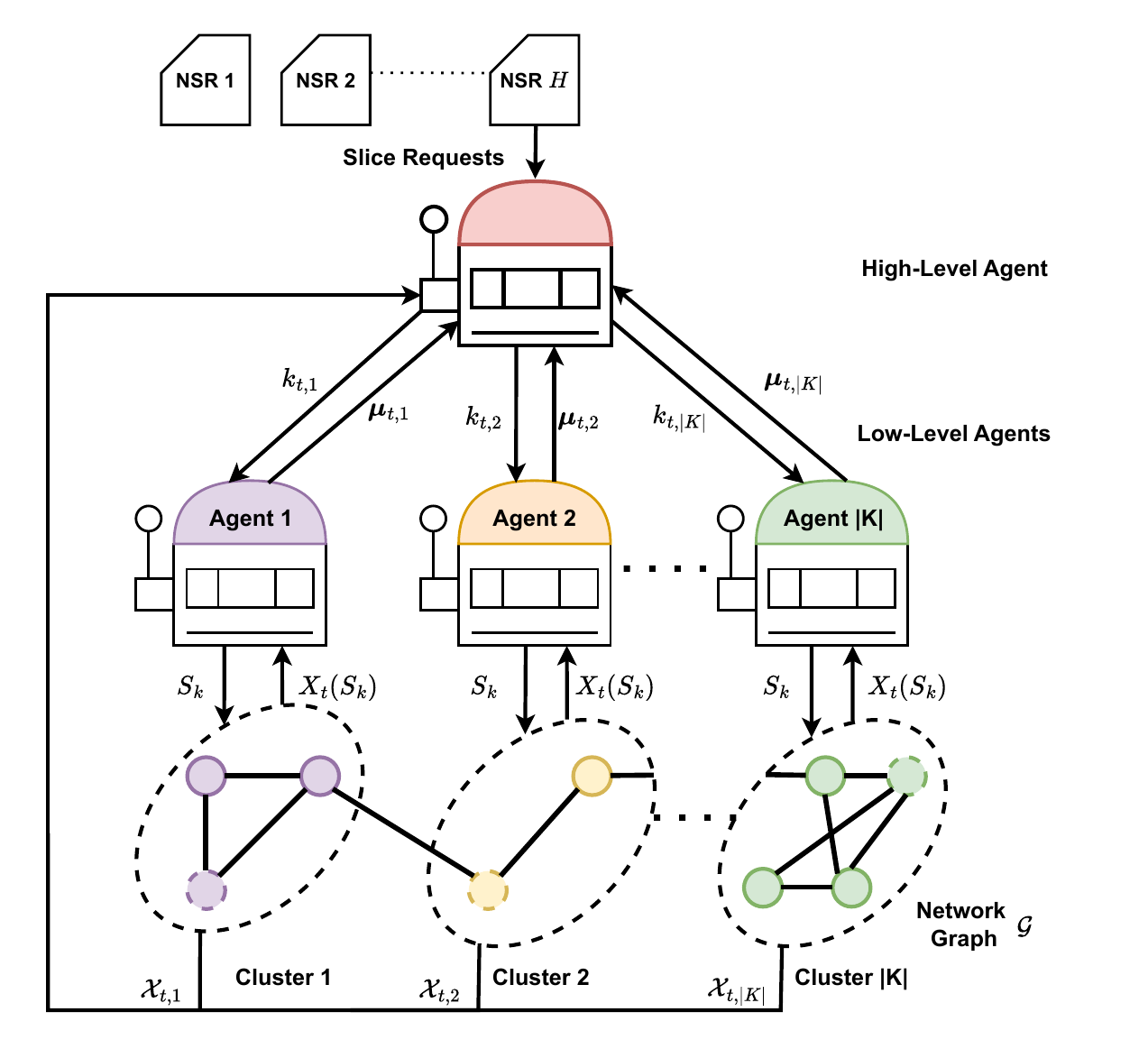}      
      \caption{\ac{helios}' Operation over a Communication Network}      
      \label{fig:masterSlave}  
    \end{figure} 

    \subsection{Background} \label{subsec:hb_bg}    
    To address the challenges mentioned in Section~\ref{sec:ProbForm}, we decompose the network slice provisioning problem into $K$ sub-problems, as seen in Fig~\ref{fig:masterSlave}.
    Each sub-problem $k \in \{1,2,...,|K|\}$ is 
    associated with a connected partition of the network graph $\mathcal{G}$.
    This decomposition leverages the structure of the physical network graph, particularly the existence of connected communities (as in~\cite{pettet2022hierarchical,10317891}), where $|K|$ is the number of communities.
    Dividing the problem in this way creates sub-problems with lower-dimensional state and action spaces, significantly reducing the overall complexity of the problem, especially for large graphs.
    We can then formulate the exploration of each sub-problem's domain hierarchically as a sequential decision problem.
    Each sub-problem $k \in K$ thus represents a sub-domain of the network slice provisioning problem that can be addressed by experts~\cite{NIPS2004_421b3ac5, hierarchicalexperts} or learned sequentially.    
    We leverage the Louvain method~\cite{Blondel_2008}, a popular complex-network algorithm which detects communities in the network by finding cuts in the network graph that can be used to partition the network into different subdomains, where each community represents a discrete region within the overall spatial area of the problem\cite{10.5555/3692070.3693934}.      
           
    In the first level of our hierarchical model, the \ac{hla} sequentially learns which discrete region (or sub-domain), managed by its respective \ac{lla}, is estimated to yield the highest multi-objective reward based on the current state.
    This is addressed using the \ac{mocmab} framework~\cite{10.1145/3394486.3403374}, to balance the trade-off between the exploration of different sub-domains and the exploitation of the sub-domains that yield the highest estimated rewards. 
    In the second-level of the model, the \ac{lla} assigned to the chosen cluster is tasked with placing the components of each \ac{nsr}    
    on the nodes in its designated sub-domain.
    The \ac{lla}'s objective is to jointly select nodes that optimize the provisioning objectives in Eq. (9), whose solution can be approximated by the \ac{comomab} framework\cite{como-ucb} to efficiently handle the exploration-exploitation trade-off for different node combinations within each \ac{lla}'s sub-domain. 
    
    \subsection{Hierarchical Multi-Armed Bandit Framework}
    We consider that the set of nodes $N$ in $\mathcal{G}$ are partitioned into $|K|$ disjoint clusters, where  $k \in \{1, 2, ..., |K|\}$.
    As depicted in Figure~\ref{fig:masterSlave}, \ac{hmab} learning consists of two parts: 
    (1) High-level policy $\pi^h$ learning at the \ac{hla}, based on the principle of contextual bandits, to orchestrate a set of high-level actions $\{a^h_k\}^K_{k=1}$, where each high-level action corresponds to selecting one of the available clusters $k$; 
    (2) low-level policy $\pi^l_k$ learning at the \acp{lla}, based on the principle of combinatorial bandits, to orchestrate a set of low-level actions $\{a^l_{S}\}$, where each \ac{lla} assigned to a cluster seeks to learn a combinatorial node selection policy that optimizes the provisioning objectives.     
    
    Consider a sequential decision-making problem over a time horizon $T$, where at each time step $t \in \{1, 2, \cdots, T\}$, the \ac{hla} observes context information $\boldsymbol{\mathcal{X}}_t \in \boldsymbol{\mathcal{X}} \subseteq [0,1]^{M|K|}$, with $||\boldsymbol{\mathcal{X}}_t||_2 \leq 1$ for all $t$.
    The context information represents the current state of the network, including the available node and link resources in each cluster $k$,
    and the aggregated resource requirements of the arriving requests.
    Specifically, we define the context $\mathcal{X}_{t,k} \in \mathbb{R}^M$ of a cluster $k$ at time slot $t$ as a vector containing $M$ relevant system information elements, such as the amount of different types of resources that are available in cluster $k$ and the requested resources by the \ac{nsr}.    
    The overall context \textcolor{black}{space} at time $t$ is therefore represented by    
    $\boldsymbol{\mathcal{X}}_t = (\mathcal{X}_{t,1}, \cdots, \mathcal{X}_{t,|K|}) \in {[0,1]^{M|K|}}$ with $\mathcal{X}_{t,k}$ being an 8-dimensional vector that is associated with each cluster $k \in K$.
    
    Based on the observed context at time step $t$, the \ac{hla} takes an action $a^h_{k} \in \boldsymbol{\mathcal{A}}^h$, where $\boldsymbol{\mathcal{A}}^h$ = $\{a_{k}^h | k \in \{1,\cdots,|K|\}\},$ and receives a corresponding multi-objective reward vector $\boldsymbol{\mu}_{t,k}$, where $\boldsymbol{\mu}_{t,k} \in [0,1]^{m+1}$.
    The selected high-level action $a^h_{k}$ directs arriving slice requests to a specific cluster, $k$. 
    Then, the \ac{lla} for that cluster selects an action $a^l_{S} \in \boldsymbol{\mathcal{A}^l_k}$, which specifies the subset of nodes ($S_k$) on which to place the requests' \acp{vnf}, where $\boldsymbol{\mathcal{A}^l_k} = \{S_k\subseteq K_k ||S_k| \leq D\},$ of which $K_k$ is the set of network nodes belonging to cluster $k$ and $D$ represents the number of \acp{vnf} to be placed.     
    Given the contextual bandit approach used by the \ac{hla}, the expected reward of selecting an action $a^h_{t,k}$ given the context $\mathcal{X}_{t,k}$ follows the linear realizability assumption and is given by: $\mathbb{E}[\boldsymbol{\mu}_{t,k}|\mathcal{X}_{t,k}, a^h_{t,k}] = \mathcal{X}_{t,k}^\top\theta^*_k,$
    where, we assume that there exists $K$ unknown weight vectors $\theta^*_1, \theta^*_2,\cdots,\theta^*_K$ for each high-level action, and where $\forall k: \theta^*_k \in \mathbb{R}^{(m+1)\times M}, ||\theta^*_k||_2 \leq 1$.
    For the sake of simplicity and without loss of generality, we adopt a shared parameter approach where $\theta^*_1 = \theta^*_2 = \cdots = \theta^*_K = \theta^*$, indicating that all arms share the same underlying unknown parameter vector rather than having disjoint arm-specific parameters.
    Based on this assumption, we define the regret of the high-level policy $\pi^h$ as
     \begin{equation}                
         R_{\pi}(T) = \mathbb{E}\left [\sum\limits_{t \in T} G(\boldsymbol{\mu}_{t,k^{*}}) - \sum\limits_{t \in T} G(\boldsymbol{\mu}_{t,k})\right]         
     \end{equation}
    where $G(\cdot)$ is an aggregation function that maps the multi-dimensional reward vector $\boldsymbol{\mu}$ to a scalar value based on the weight vector $\mathbf{w}$ (Eq. (9)), $k^{*}$ is the index of the best high-level action at time step $t$ and $k$ indicates the index of the high-level action selected by the \ac{hla}. 
    In the formulated \ac{hmab}, the goal is to learn a cluster selection policy at the \ac{hla} and a combinatorial node selection policy at the \ac{lla}.             
    To learn a cluster selection policy, we consider the following approach:
    
    \subsubsection{Multi-Objective Cluster Selection}   
    Let us define the known mean reward vectors of each cluster $k\in\{1,\ldots,|K|\}$ as $\boldsymbol{\mu}_k$. 
    We consider that for a given aggregation function $G(\boldsymbol{\mu}_k)$, the optimal cluster selection policy seeks to find a strategy such that the index of $G(\bar{\boldsymbol{\mu}_k})$ is as large as possible, where $\bar{\boldsymbol{\mu}}_k = \frac{1}{T} \sum^T_{t=1}\boldsymbol{\mu}_{t,k}$.
    This ensures that clusters are selected with the aim of maximizing the aggregation function, towards the optimization of the provisioning objectives.
    In this work, we consider that the aggregation function that scalarizes inputs from different objectives, is the \ac{ggf}~\cite{weymark1981generalized}.
    The \ac{ggf} is a non-linear, concave function, and is a special case of the \ac{owa} aggregation operators~\cite{yager1988ordered} that seek to preserve impartiality with respect to individual objectives.        
    Rather than using a deterministic strategy that always selects the single cluster maximizing the aggregated reward $G$, we adopt a mixed strategy that optimizes a probability distribution $\boldsymbol{\alpha}_t \in \mathbb{A}$ over all clusters, from which a cluster is sampled.    
    Specifically, the probability distribution is the simplex $\mathbb{A} = \{ \boldsymbol{\alpha}=(\alpha_1,\ldots,\alpha_{|K|}) \in [0,1]^{|K|} : \lVert\boldsymbol{\alpha}\lVert_1 = 1 \}$, according to which a cluster $\alpha_k$ is selected.    
    As explained in \cite{10.5555/3305381.3305446}, the optimal mixed strategy cluster selection policy is given by solving the following optimization problem, $\textcolor{black}{\boldsymbol{\alpha}^*\in\underset{\boldsymbol{\alpha}\in \mathbb{A}}{\arg\max}\quad G\left(\sum_{k=1}^{|K|} \alpha_k \boldsymbol{\mu}_{k}\right)}$, which determines a probability distribution over clusters that maximizes the \ac{ggi} of the expected aggregated rewards.    
    However, based on the probability distribution used by the mixed strategy, clusters with smaller estimated rewards are periodically selected to balance the trade-off between exploration of potentially useful clusters that could lead to good multi-objective rewards and exploitation of cluster that are known to have high estimated rewards.    
    By leveraging the \ac{ggf} as the aggregation function in the optimal cluster selection strategy, the regret of the high-level policy can be estimated as: 
    \begin{multline}
        \hat{R}_\pi{(T)} = G\left(\frac{1}{T}\sum_{t=1}^{T}\sum_{k=1}^{K}\alpha_{t,k}^{*}\boldsymbol{\mu}_{t,k^{*}}\right) - \\
        G\left(\frac{1}{T}\sum_{t=1}^{T}\sum_{k=1}^{K}\alpha_{t,k}\boldsymbol{\mu}_{t,k}\right)
    \end{multline}

    Given this definition, the goal of the high-level policy is to minimize the regret by maximizing the \ac{ggf} of the aggregated multi-objective reward (second term in Eq. (11)), based on the current parameter estimates of the cluster selection strategy.    
    Specifically, we utilize ridge regression~\cite{cessie1992ridge}  to get the expected reward $\boldsymbol{\mu}_{t,k}$ for a particular action $a^h_{t,k}$ at time $t$, given the context $\mathcal{X}_{t,k}$. 
    The parameter estimate for objective $o$ is given by $\hat{\boldsymbol{\theta}}^o_{t} = (A_t + \gamma \textbf{I}_M)^{-1} \mathbf{b}^o_{t}$.
    Here, $A_t = \sum^{t-1}_{\tau=1} (\mathcal{X}_{\tau,k_{\tau}} \mathcal{X}_{\tau,k_{\tau}}^\top)$
    accumulates the outer products of contexts from previously selected clusters, and     
     $\mathbf{b}^o_{t} = \sum^{t-1}_{\tau=1}\mathcal{X}_{\tau,k_{\tau}} r_\tau^{o}$ 
    accumulates the products of contexts with the observed rewards, 
    where $r_\tau^{o} \in \mathbf{r}_{\tau}$ is the reward of objective $o$ observed from the combinatorial action of the \ac{lla} in cluster $k_{\tau}$ at time $\tau$ and $\mathbf{r}_{\tau} = {\boldsymbol{X}}_{\tau}(S_{k_{\tau}})$.
   
    The cluster selection policy $\alpha_t$ seeks to maximize the \ac{ggf} of the expected aggregated multi-objective rewards through \ac{oga} by optimizing the function $G\left(\sum^{|K|}_{k=1}\alpha_{t,k}\hat{\boldsymbol{\mu}}_{t,k}\right)$. 
    Starting from a uniform distribution $\alpha_t = (1/|K|,...,1/|K|)$, we perform $Z$ gradient ascent steps, where at each step $z$, we perform the following:
    \begin{itemize}
        \item \textcolor{black}{Compute the gradient}: $\mathbf{g}^{(z)} = \nabla_{\boldsymbol{\boldsymbol{\mu}}} G\left(\sum_{k=1}^{K} \alpha_k^{(z)} \hat{\boldsymbol{\mu}}_{t,k}\right) \cdot \hat{\mathbf{M}}_t$, where $\hat{\mathbf{M}}_t = [\hat{\boldsymbol{\mu}}_{t,1}, \ldots, \hat{\boldsymbol{\mu}}_{t,|K|}]$ is the matrix of estimated mean reward vectors of each cluster $k$.
        \item Next, we take a gradient step: $\tilde{\boldsymbol{\alpha}}^{(z+1)} = \boldsymbol{\alpha}^{(z)} + \eta_z \mathbf{g}^{(z)}$, with step-size $\eta_z = \frac{1}{\sqrt{z+1}}$
        \item Finally, this is projected onto the simplex: $\boldsymbol{\alpha}^{(z+1)} = \Pi_{\mathbb{A}}\left(\tilde{\boldsymbol{\alpha}}^{(z+1)}\right)$.        
    \end{itemize}

    After $Z$ iterations, we use $\boldsymbol{\alpha}_t = \boldsymbol{\alpha}^{(Z)}$ as our mixed strategy and sample cluster $k_t$ according to this distribution.
    This approach ensures convergence to a local optimum of the \ac{ggi} while maintaining the constraint that $\boldsymbol{\alpha}_t$ remains a valid probability distribution~\cite{10.1145/3394486.3403374}.

    \subsubsection{Combinatorial Node Selection} 
    Upon receiving an \ac{nsr} $\mathcal{H}_t$ directed by the \ac{hla} to a cluster $k_t$, the designated \ac{lla} for that cluster is tasked with the goal step of selecting an optimal set of physical nodes to host the $D$ \acp{vnf} comprising the \ac{sfc}. 
    This selection is formulated as a \ac{comomab} problem~\cite{como-ucb}, where the \ac{lla} must learn to make choices that align with the overall system objectives defined in Eq~(9).   
    To formulate the problem, we define $K_k$ as the set of network nodes belonging to cluster $k$.
    Each network node in set $K_k$ can be selected by the \ac{lla} for deploying a \ac{vnf}, therefore each network node in $K_k$ is a \textit{base arm} of the agent.
    We define the \textit{power set} $\mathcal{P}(K_k)$ of set $K_k$ as the set of all possible combinations of base arms $\in K_K$, i.e., $\mathcal{P}(K_k) = \{ S_k| S_k\subseteq K_k \}$. We now define a \textit{super arm} $S_k$ for cluster $k$ as a subset of the set $K_k$ of base arms, which represents the network nodes on which the \ac{sfc}'s \acp{vnf} will be simultaneously deployed, i.e., $S_k\in\mathcal{P}(K_k)$. It is worth noting that $|\mathcal{P}(K_k)|=2^{|K_k|}$, which makes the problem NP-hard as it scales exponentially with the $k$-th cluster size $|K_k|$.

    At each time step $t$, the \ac{lla} pulls the super arm $S_t$ and receives a reward
    $\boldsymbol{X}_t(S_k) = (X_{t,1}, \ldots, X_{t,|S_k|}) \in [0, 1]^{|S_k|\times (m+1)}$ 
    , and the outcomes of the base arms in $\boldsymbol{X}_t$ are assumed to be independent.
    The rewards represent the acceptance of the \acp{vnf} in the selected, as well as the resource utilization on the selected nodes as a result of deploying the request.
    The final goal of the \acp{lla} is to eventually learn the optimal super arm $S_k^*$ (i.e., set of nodes in $K_k$) that optimizes the objectives defined in Eq~(8).
    The goal of the \ac{lla} is to learn the optimal super arm    
    $S_{kt}^* \in \mathcal{P}(K_k)$
    , over time.    
    However, selecting the optimal super arm, which is given by:    
    \begin{equation}
        S^*_{kt} = \underset{S\in \mathcal{P}(K_k) }{\arg\max} \lVert X_t(S) \rVert
    \end{equation}

    is an NP-hard problem~\cite{liu2024learning, ijcai2024p44}.    
    Hence, the \acp{lla} in the \ac{hb} framework uses the \ac{ucb} to learn the rewards of the different arms in a given super arm and picks the super arm with the higher upper confidence bound.
    More specifically, a simplified version of the \ac{comoucb}~\cite{como-ucb} algorithm is developed to address the exploration and exploitation dilemma of the considered problem.

    \subsection{Proposed Solution}    
    \begin{algorithm}
    \footnotesize
    \DontPrintSemicolon
    
    \SetKwFor{Loop}{Loop}{}{}
    \SetKwFunction{HighLevelAgent}{HighLevelAgent}
    \SetKwFunction{LowLevelAgent}{LowLevelAgent}
    \SetKwFunction{PullArm}{PullArm}
    \SetKwFunction{WaitNSRs}{WaitNSRs}
    \SetKwFunction{Louvain}{Louvain}
    \SetKwFunction{GetClusterContexts}{GetClusterContexts}
    
    \textbf{Input:} Regularization $\gamma > 0$, learning rate $\eta_z > 0$, number of gradient steps $Z$ \\
    \textbf{Initialize:} $A_t \gets \gamma I_M$, $\mathbf{b}_t^{o} \gets \mathbf{0}_M$ $\forall o \in [m+1]$\\
    $\text{UCB}^o_{i}(t) \gets~+\infty, \forall o \leq [m+1], \forall i\in[K_k], \forall t\in [\mathcal{H}_t]$
    
    \tcp{Create Set of Clusters based on $\mathcal{G}$}
    $|K| \gets $\Louvain{$\mathcal{G}$}\;        
    $\boldsymbol{\alpha}_t \gets (1/|K|,...,1/|K|)$    
    
    \For{$t \in \{1,\ldots, T\}$}{
            \tcp{Collect NSRs arriving during timeslot $t$}        
        $\mathcal{H}_t \gets \WaitNSRs(t)$\;
        $k_t \gets$ \HighLevelAgent{$\mathcal{H}_t$} \;
        $\mathbf{X}_{t}(S_{k_t}) \gets$ \LowLevelAgent{$\mathcal{H}_t,k_t$} \;
        $\mathbf{r}_t = \mathbf{X}_t(S_{k_t}) \in \mathbb{R}^{m+1}$\;

        \tcp{Regression Update}              
        $A_{t} \gets A_t + \mathcal{X}_{t,k_{t}}\mathcal{X}_{t,k_{t}}^T$ \;                       
        
        \For{$o \in [m+1]$}{            
            $b^o_{t+1} \gets b^o_{t} + \mathcal{X}_{t,k}  r_t^{o}$
        } 
        }

    \hrule

    \SetKwFunction{FHighLevelAgent}{HighLevelAgent}
    \SetKwProg{Fn}{Function}{:}{}
    \Fn{\FHighLevelAgent{$\mathcal{H}_t$}}{        
            \tcp{Observe Cluster Contexts at timeslot $t$}
            $\boldsymbol{\mathcal{X}}_t \gets $ \GetClusterContexts($t$)
    
            \For{$o \in [m+1]$}
            {
                $\hat{\boldsymbol{\theta}}_t^{o} \gets A_t^{-1} \mathbf{b}_t^{o}$\; 

                \For{$k \in K$}{
                    $\hat{\mu}_{t,k}^{(o)} \gets \mathcal{X}_{t,k}^\top \hat{\boldsymbol{\theta}}_t^{o}$\; 
                }
            }
            \tcp{Perform Online Gradient Ascent to optimize mixed strategy}
            $\boldsymbol{\alpha}^{(0)} \gets \boldsymbol{\alpha}_t$ 
        
            \For{$z \in [Z]$}{
                $\nabla_{\boldsymbol{\mu}} G\left( \sum_{k=1}^K \alpha_k^{(z-1)} \hat{\boldsymbol{\mu}}_{t,k} \right)$\; 
                $\boldsymbol{\alpha}^{(z)} \gets \Pi_{\mathbb{A}} \left( \boldsymbol{\alpha}^{(z-1)} + \eta_z \cdot \nabla \right)$ 
            }
        
            $\boldsymbol{\alpha}_t \gets \boldsymbol{\alpha}^{(Z)}$\; 
            $k_t \sim \text{Categorical}(\boldsymbol{\alpha}_t)$\; 

            \tcp{Send request to LLA of cluster $k_t$}           
            \KwRet $k_t$ \;
    }

    \hrule

    \SetKwFunction{FLowLevelAgent}{LowLevelAgent}
    \SetKwProg{Fn}{Function}{:}{}
    \Fn{\FLowLevelAgent{$\mathcal{H}_t, k_t$}}{        
        \For{$t \in  \{1, \ldots, |\mathcal{H}_t|\}$}{        
            \For{$i \in [K_k], o \in [m+1]$}{
                \tcp{Update Upper Confidence Bound}
                \If{$P_i> 0$}{ 
                    $\text{UCB}^o_{i}(t) \gets \bar{X}^o_{i}(t-1) + \sqrt{\frac{3\log t}{2P_i}}$ \label{alg:moonline:ucb_update}
                }
            }
            
            $S_t \gets \operatorname*{arg\,max}_{S \in \mathcal{S}} \sum_{i \in S} \sum_{o=1}^{m+1} \text{UCB}^o_{i}(t)$ \;
            
            \tcp{Play super arm $S_t$ and observe rewards}

            $\boldsymbol{X}_t (S_k) = (X_{t,1}, \ldots, X_{t,|S_t|}) \gets $ \PullArm($S_t$) \;
    
            \For{$i \in S_k$}
            {
                \tcp{Increment pull counter for arm $i$}
                $P_{i+1} \gets P_i + 1$\;                 
                \For{$o \in [m+1]$}{
                    $\bar{X}^o_{i}(t) \gets \frac{P_i\bar{X}^o_{i}(t-1) + X^o_{i}(t)}{P_{i+1}}$
                }
            }
        } 
        \KwRet $\boldsymbol{X}_{t}(S_k)$ \;
    }
    \caption{\ac{helios}}
    \label{alg:moonline}
    \end{algorithm}
    
    Algorithm~\ref{alg:moonline} works as follows.
    First (lines 2 and 3), we initialize the contextual and combinatorial bandit parameters $A_t$, $\mathbf{b}_t$, and $UCB^o_{i}$ which represent the identity matrix, scaled by $\gamma$ for ridge regression, the reward accumulation vector for each objective $o$, and the upper confidence bound of the reward for each base super arm, respectively.
    Then (line 4), we receive the $|K|$ clusters based on the communities detected on the network graph $\mathcal{G}$.
    In line 5, we initialize a (uniform) probability distribution over the clusters $\boldsymbol{\alpha_{t}}$ which determines the initial probability of selecting each cluster..
    In lines 6 and 7, we begin by collecting the \acp{nsr} arriving at the current time slot.         
    Based on the collected \acp{nsr} and the current resources in the clusters, the \ac{hla} estimates the reward of placing the requests in each cluster (lines 15-19).    
    In lines 20-25, we update $\boldsymbol{\alpha_{t}}$ using projected gradient ascent to maximize the \ac{ggi}-aggregated expected reward, and then sample a cluster according to the resulting distribution.    
    
    Based on the selected cluster in the \ac{hla} procedure, the \ac{lla} function determines the combination of nodes within the cluster to deploy the \acp{vnf} of a request.
    For each request in the current time slot $\mathcal{H}_t$, the \ac{lla} looks to improve its \ac{vnf} placement strategy through exploration and exploitation.
    In lines 29-31, for each node $i$ in the selected cluster $K_k$ and each objective $o$, the \ac{lla} evaluates the potential rewards from selecting that node based on the empirical mean rewards from previous times a node was selected.     
    From lines 32-33, the \ac{lla}, based on the \ac{comoucb} algorithm~\cite{como-ucb}, selects the super arm $S_t$ (i.e., set of nodes) that maximizes the sum of \ac{ucb} values across all the selected nodes and objectives. 
    If the \acp{vnf} of the request are deployed on the set of nodes, then the \ac{lla} receives a multi-dimensional reward $\boldsymbol{X_t}(S_k)$ that captures the provisioning objectives (i.e., the acceptance of the request and the utilization of resources in the cluster) (lines 34-37).
    Based on the selected nodes by the \ac{lla}, the shortest path between the nodes is computed using Dijkstra’s algorithm.
    In lines 11-13, we update the parameters of the contextual bandit algorithm based on the multi-dimensional rewards of the \ac{lla}'s selected nodes, in order to improve the overall learning procedure.

   \section{Performance Evaluation} \label{sec:perfeval}      
   \subsection{Simulation Settings}\label{subsec:SimSetup}
    We evaluated the performance of our proposed approach against five baselines on two real-world network topologies of different size, namely: GEANT (22 nodes, 33 links, which is a pan-European research and education data network)~\cite{6027859} and DTelekom - DT2 - (68 nodes, 272 links) which is a sample topology of Deutsche Telekom~\cite{rossi2011caching}.
    The simulations, proposed approach, and baseline algorithms were developed and implemented in Python using NetworkX to simulate different network scenarios on the two network topologies.    
    We select the Louvain community detection algorithm to identify the $|K|$ communities (or clusters) within the network topology based on the topological features of each generated network graph.
    One \ac{lla} agent is deployed within each cluster to handle the joint placement of \acp{vnf} on the devices within the cluster.    
    The CPU, RAM, storage, and GPU capacities of network nodes $\varphi^m_v$, were uniformly generated within the range of $[5, 1000]$ resource units, while the bandwidth of each link is in the range $[50, 5000] \unit{\mega\bit\per\second}$.       
    For each \ac{nsr} we randomly generate an \ac{sfc} with $\{2, 3, 4\}$ \acp{vnf}, where each \ac{vnf} represents a generic \ac{nf}.
    The resource requirements of each \ac{vnf} and virtual link in a request are uniformly generated within the range of $[5, 50]$ resource units and $[50, 100] \unit{\mega\bit\per\second}$, respectively.     
    
    To conduct realistic evaluations, we evaluate the performance of our algorithm in an online setting with $T =$ 5000 time slots.     
    We assume a random number $v_t$ of \acp{nsr} arriving in the system at the beginning of each time slot, which defines the time between successive requests.    
    We model $v_t$ as a Poisson process in which all $v_t, \forall t \in [T]$ follow a Poisson distribution, $\text{Pois}(\lambda)$, with identical arrival rate $\lambda$, where $\lambda$ is the average number of arrivals per slot.    
    We evaluated scenarios, where $\lambda = 2$ and $\lambda = 5$.
    The lifetimes $\delta_h$ of the requests (in minutes) are generated from a uniform distribution $\delta_h\sim\mathcal{U}(\{10, \zeta\})$, where $\zeta$ is the upper bound of the request duration and we evaluated the scenario in which $\zeta = 50$ to investigate the impact of different request lifetimes on the system.    
    \textcolor{black}{Finally, we initialize the weight vector $\mathbf{w}$ to $w_{o} = 2^{-o+1}, o \in [m+1]$~\cite{10.5555/3305381.3305446}.}
    Table~\ref{tab:simParams} summarizes the simulation parameters used.
    
    \begin{table}[ht!]
        \centering
        \begin{tabular}{p{5.8cm}p{2.2cm}}\toprule
            Parameters & Values\\
            \midrule
            Clusters in GEANT and DTelekom, $|K|$ & \{3, 5\}\\   \hline       
            \textcolor{black}{GGI Weights} $G_\mathbf{w}$ & \textcolor{black}{$w_{o} = 2^{-o+1}$}\\   \hline       
            \ac{oga} Iterations $Z$ & $10$ \\\hline
            Learning Rate $\eta_z$ & $\frac{1}{\sqrt{z+1}}$ \\ \hline
            Regularization Parameter $\gamma$ & $0.1$ \\\hline            
            \ac{vnf} Resource Requirements  & $[5, 50]$ units\\\hline
            Node Capacity $\varphi^m_v$ & $[5, 1000] $ units\\\hline
            Virtual Link Requirements & $[50, 100]\unit{\mega\bit\per\second}$ \\ \hline
            Link Capacity & $[500, 5000]\unit{\mega\bit\per\second}$\\\hline
            \ac{sfc} Length (\acp{vnf} per request) $\psi$ & \{2, 3, 4\}\\
            \bottomrule
        \end{tabular}        
        \caption{Simulation Parameter Settings}        
        \label{tab:simParams}
    \end{table}
    
    \subsubsection{Baselines}
   
    We compare \ac{helios} against five centralized baselines: 
   
    \begin{itemize}
       \item \textit{Random}: The random baseline is a myopic policy that uniformly and jointly assigns \acp{vnf} of an \ac{sfc} to different nodes in the network without considering any context information. 
       \item \textit{$\epsilon$-greedy}: This approach uses an $\epsilon-$greedy strategy to determine where to deploy the \acp{vnf} of a request, based on exploration parameter, $\epsilon$, which we set to $\epsilon= 0.5$ in our evaluations.       
       \textcolor{black}{\item \textit{\ac{linucb}}~\cite{li2010contextual}: \acs{linucb} is a learning-based algorithm that leverages context information to determine where to place requests given the entire topology.}
       \textcolor{black}{\item \textit{\ac{c2ucb}}~\cite{qin2014contextual}: The \ac{c2ucb} algorithm combines context information with combinatorial node selection to determine the nodes on which to place a request jointly.} 
       \textcolor{black}{\item \textit{\ac{cts}}~\cite{pmlr-v80-wang18a}: \acs{cts} is a lightweight solutions that learns a provisioning policy based on estimating the posterior distribution of the expected rewards of different placements conditioned on the context information.} 
    \end{itemize}    
        
    \subsection{Simulation Results} \label{subsec:Results}    
    We evaluated the performance of our approach by considering the following three metrics: Average Acceptance Rate and Average CPU Resource Utilization and Average Execution Time.
    For reliable simulations, we conduct 30 simulation runs and present the average performance.    

    \begin{figure*}
    \centering
        \begin{subfigure}[t]{0.40\textwidth}
            \centering
            \includegraphics[ width=\textwidth]{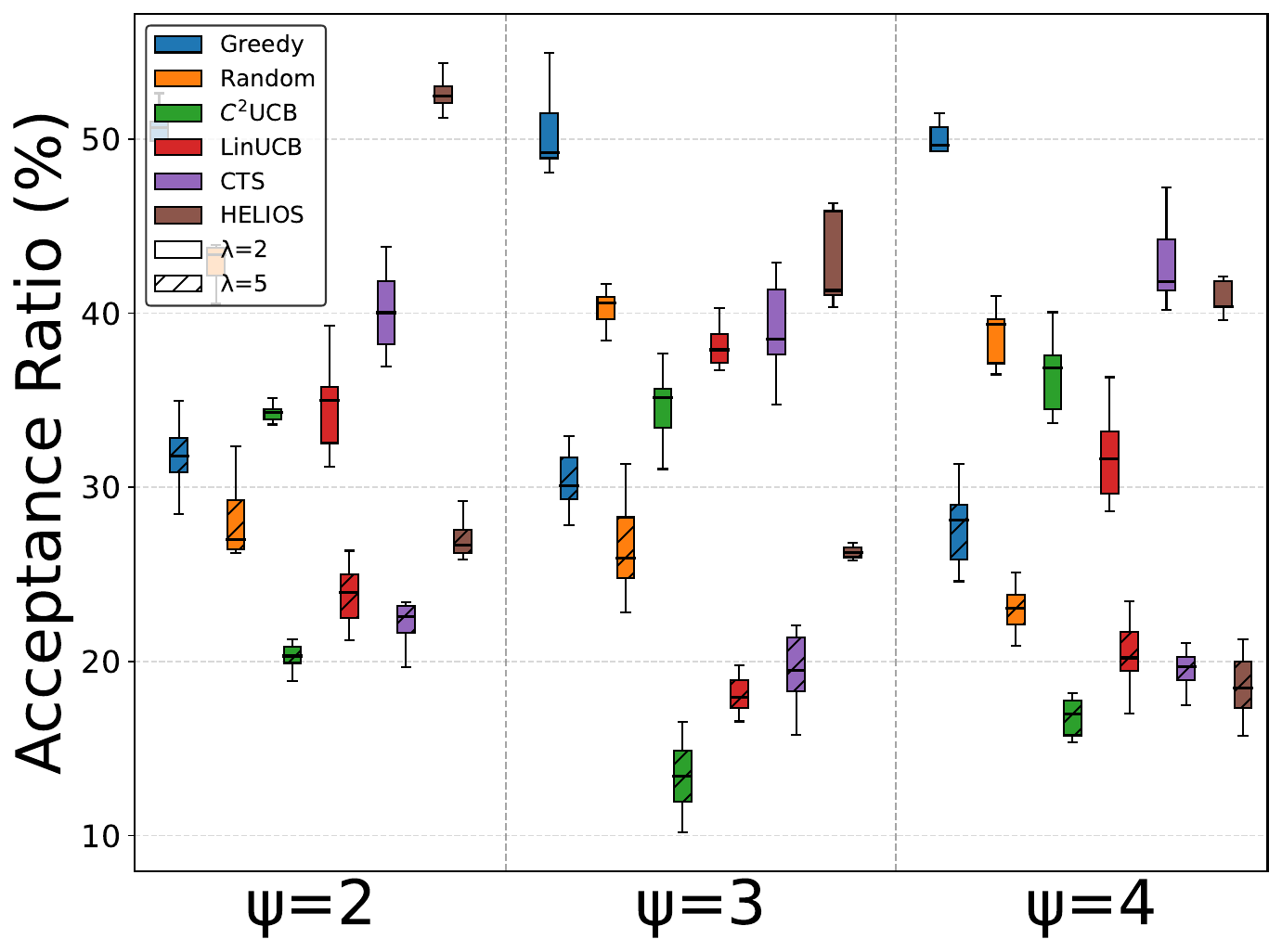}\hfill                      
            \caption{Average Acceptance Ratio}
            \label{fig:acc_ratio_GEANT}
        \end{subfigure}~
        \begin{subfigure}[t]{0.41\textwidth}
            \centering
            \includegraphics[width=\textwidth]{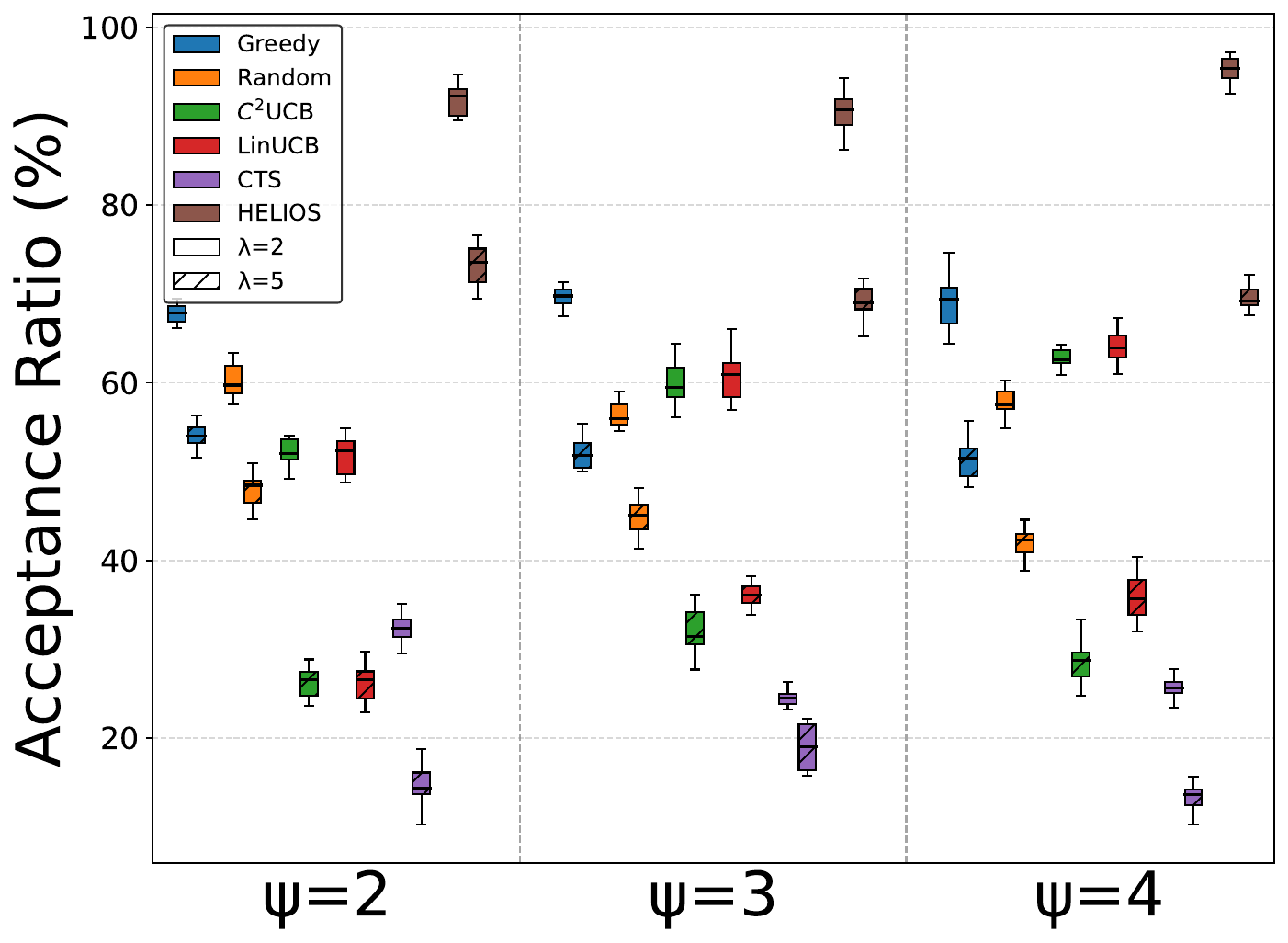}\hfill                      
            \caption{Average Acceptance Ratio}
            \label{fig:acc_ratio_DT2}
        \end{subfigure}~
    \caption{Acceptance Ratio Performance over the a.) GEANT and b.) DT2 Network Topologies}    
    \label{fig:acceptanceRatio}
    \end{figure*}
    
    \begin{figure*}
    \centering
        \begin{subfigure}[t]{0.24\textwidth}
            \centering            
            \includegraphics[width=\textwidth]{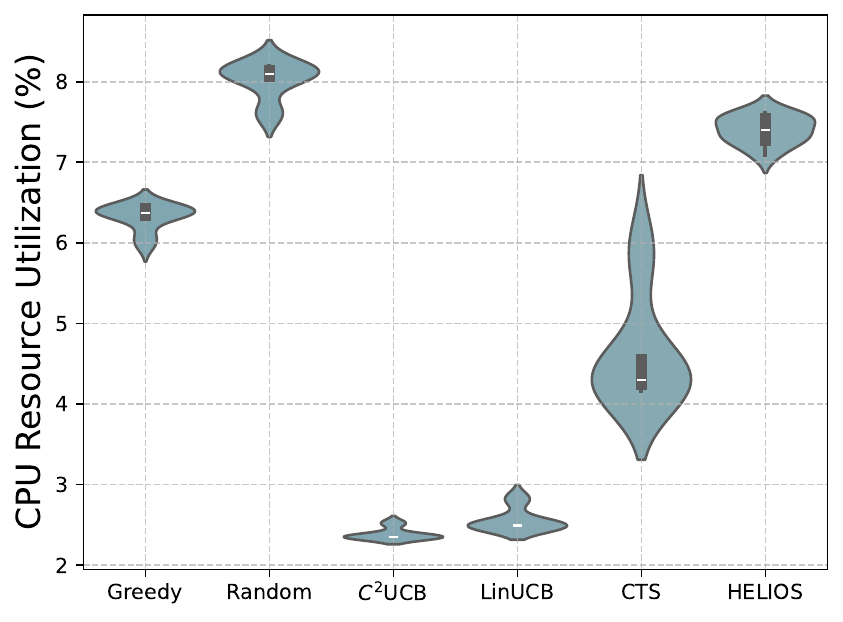}\hfill                      
            \caption{Average Resource Utilization}
            \label{fig:res_util_GEANT}
        \end{subfigure}~        
        \begin{subfigure}[t]{0.24\textwidth}
            \includegraphics[width=\textwidth]{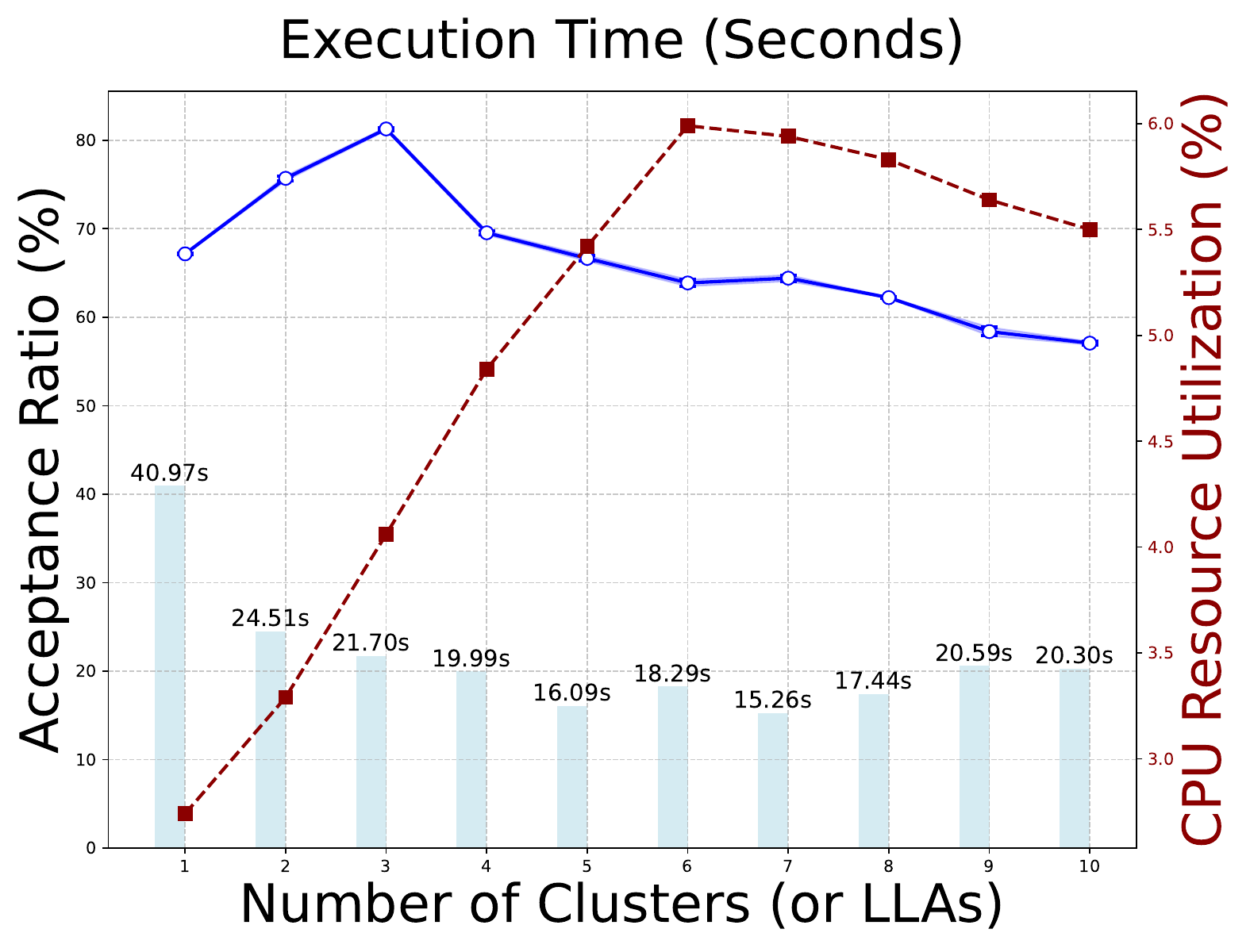}\hfill                    
            \caption{Cluster-based Performance}
            \label{fig:num_clusters_GEANT}
        \end{subfigure}~
        \begin{subfigure}[t]{0.24\textwidth}
            \centering
            \includegraphics[width=\textwidth]{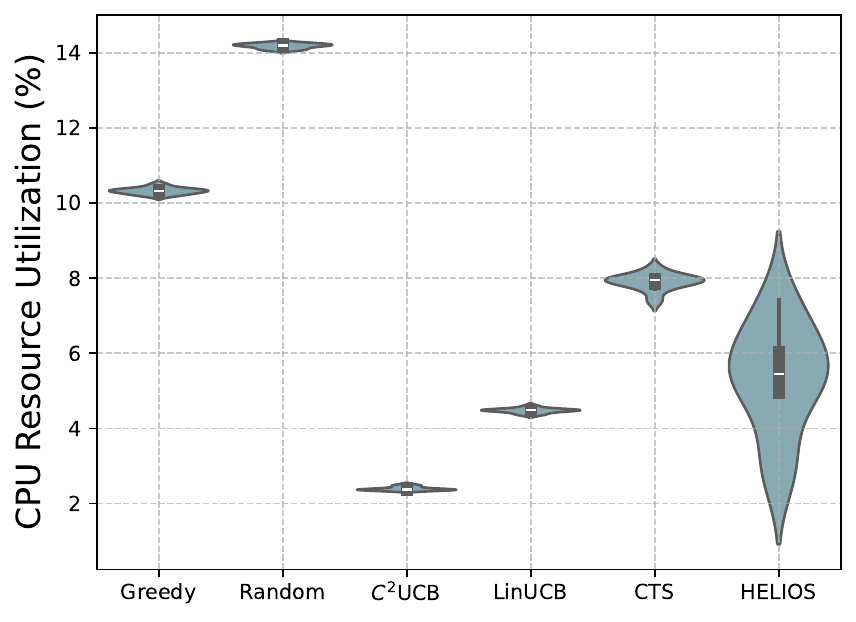}\hfill                      
            \caption{Average Resource Utilization}
            \label{fig:res_util_DT2}
        \end{subfigure}~
        \begin{subfigure}[t]{0.24\textwidth}
            \centering
            \includegraphics[width=\textwidth]{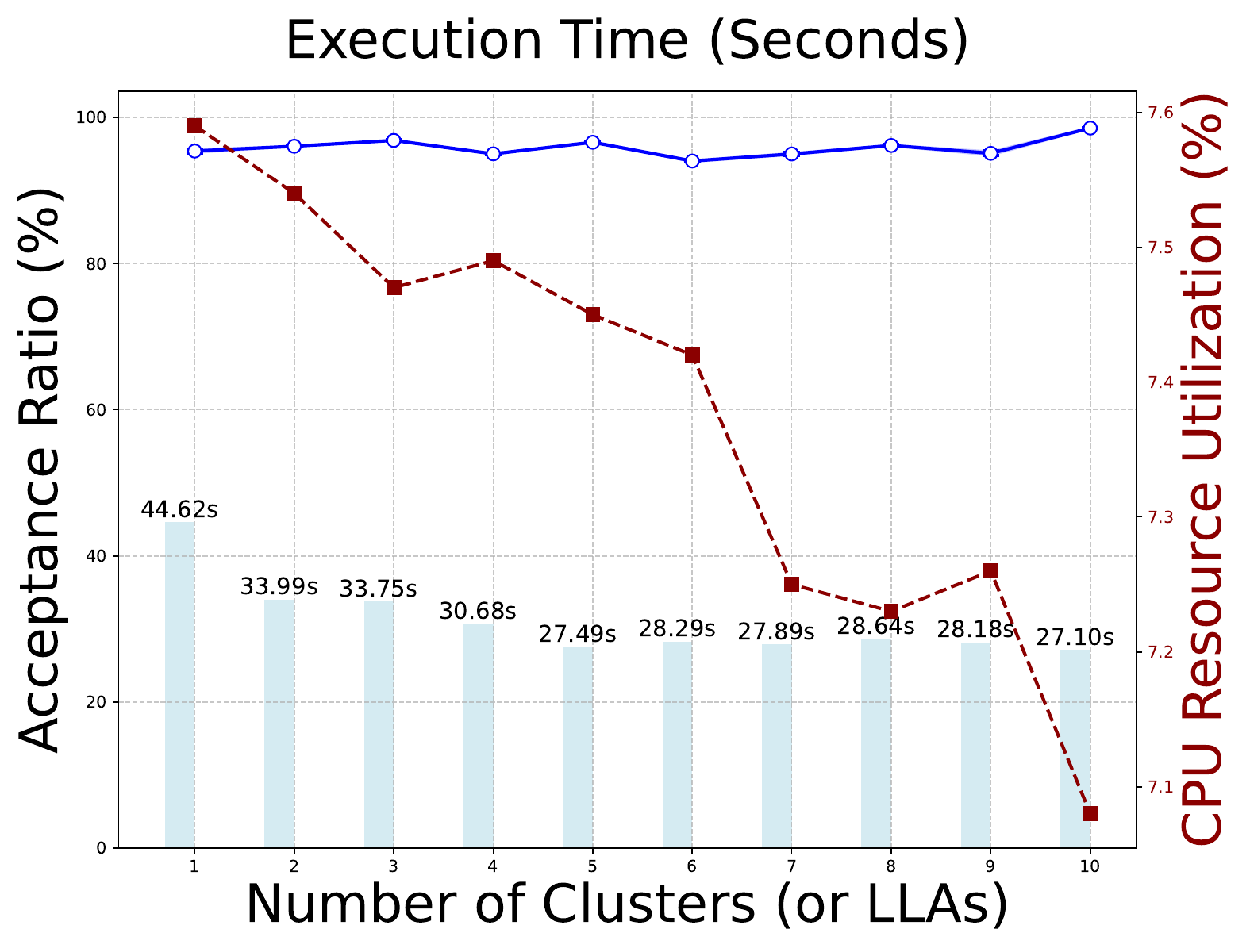}                                 
            \caption{Cluster-based Performance}
            \label{fig:num_clusters_DT2}
        \end{subfigure}
    \caption{Average Resource Utilization and Cluster-based Performance over the a.)-b) GEANT and c.)-d.) DT2 Network Topologies}    
    \label{fig:resourceUtilization}
    \end{figure*}

    \subsubsection{Request Acceptance Rate} \label{subsubsect:acceptance}
    is defined as the ratio between the total number of \acp{nsr} received by the \ac{inp} and the total number of \acp{nsr} accepted onto the infrastructure.
    Fig.~\ref{fig:acc_ratio_GEANT} and Fig.~\ref{fig:acc_ratio_DT2} show the performance of our approach compared to the baselines on the GEANT and DT2 topologies, respectively. 
    We see that on the larger DT2 topology, our solution is able to admit a higher number of \acp{nsr} (up to 97\%) under different arrival conditions ($\lambda$) and across the different chain lengths ($\psi$), compared to the baselines.    
    However, in the smaller GEANT topology, the overall performance decreases across the different scenarios. 
    This is attributed to the fact that there are fewer nodes per cluster in the GEANT topology, making it less likely for each cluster to contain an optimal combination of nodes that can meet the provisioning objectives. 
    This is especially true for scenarios with longer chains and higher arrival rates, as node resources are more likely to be congested, leading to a lower overall acceptance rate.     
    Despite this, our results suggest that \ac{helios} can learn a hierarchical provisioning policy that leads to better performance by fragmenting the original problem into sub-domains due to the inherent spatial structure of communication networks and leveraging context information in the placement decision.    
    
    \subsubsection{Average Node Resource Utilization} \label{subsubsect:utilization}
    In Fig.~\ref{fig:res_util_GEANT} and Fig.~\ref{fig:res_util_DT2}, we observe the average CPU resource utilization per node in the considered system. 
    Based on the provisioning objectives, we see that the provisioning policy learned by our approach leads to an average resource utilization per node per time slot that is relatively low, approximately 5\%, especially in the larger DTelekom network, which is reflected in the higher acceptance ratio for the network scenarios considered in Fig. ~\ref{fig:acc_ratio_DT2}. 
    However, in the smaller GEANT topology, the average CPU resource utilization is considerably higher reaching up to 9\% average utilization, which leads to a lower acceptance ratio for this topology, as fewer requests are deployed on the GEANT network. 
    This highlights how our proposed solution is able to minimize average resource utilization, while seeking to maximize the number of accepted requests, by learning a cluster selection policy and a combinatorial node selection policy that achieves both objectives.
    Compared to the \ac{ucb}-based approach employed by \ac{helios} and other baselines, the \ac{cts} approach typically leads to an unbalanced distribution of actions due to the random nature of the sampling process from the posterior distribution, which contrasts the more consistent policy of \acp{ucb}-based solutions~\cite{network2030023}.
    \subsubsection{Average Execution Time} \label{subsubsect:executiontime}    
    \begin{table}[ht!]
            \centering
            \caption{Avg. Execution Times of Different Algorithms [\si{\second}]}        
            \begin{tabular}{ccccc} \toprule
                 Topology & C$^2$UCB &\acs{linucb} & \acs{cts} & \ac{helios} \\ \midrule
                 GEANT & 26.7 & 51.3 & 29.8 & 38.4\\ \hline
                 DTelekom & 71.2 & 182.3 & 46.4 & 96.5\\                       
                 \bottomrule
            \end{tabular}        
            \label{tab:executions}
    \end{table}
    We evaluate the execution times of the different learning-based baselines on deploying a set of requests ($\sim$25000) across the considered topologies.
    As the random and $\epsilon-$greedy approaches are heuristically simple and, therefore, have negligible execution times, we omit their results due to brevity.    
    Our results are shown in Table~\ref{tab:executions}.
    We see that on average, our proposed solution had a higher execution time compared to the majority of baseline solutions on the GEANT and DT2 topologies, taking 38 s and 96 s, respectively.
    This is a result of our our hierarchical approach which adds logical complexity and requires multi-level decisions on the placement decision of requests.    
    Furthermore, the \ac{hla} takes multiple \ac{oga} steps during the \ac{lla} selection, which adds to the longer execution time.

    \subsubsection{Cluster-based Performance}
    Fig.~\ref{fig:num_clusters_GEANT} and Fig.~\ref{fig:num_clusters_DT2} show the scalable performance of our solution for different numbers of clusters.
    As it can be seen in the results, the performance of our solution begins to decrease after dividing the topology into more than three sub-domains, or clusters, in the GEANT topology.
    However, it does not show a similar trend in the larger DT2 topology. 
    This is attributed to the average size of clusters in each topology, as more clusters in the GEANT topology reduces the number of nodes per cluster affecting the performance of the \acp{lla}, while the number of nodes per cluster is larger in DT2.
    We also see that increasing the number of clusters in each topology reduces the average execution time of our solution, highlighting the tradeoff between performance and speed on smaller topologies.   
    
    \section{Conclusion} \label{sec:conclusion}
    We introduced a hierarchical placement learning solution to address the network slice provisioning problem in edge networks.
    Our approach uses a hierarchical bandit approach to learn local provisioning strategies over subsets of the network topology by leveraging the inherent spatial structure of communication networks.    
    In particular, our solution partitions the overall network topology into clusters, where we address the network slice provisioning problem in each cluster as individual bandit problems.
    Each \ac{lla} learns slice provisioning strategies for its local topology, while the \ac{hla} coordinates resource allocation across sub-domains to satisfy the provisioning objectives.    
    We show that compared to myopic and centralized online learning solutions, our proposed approach is able to outperform them for the considered metrics, but is constrained by the quality of the formulated clusters.
    The integration of an approach that can recursively and adaptively partitioning the network graph to improve the performance and robustness of our solution on various types of graphs (as in \cite{li2025navigatinghighdimensionalsearchspace}), is left to future work.
    
\bibliographystyle{IEEEtran}
\bibliography{IEEEabrv,references.bib}

\end{document}